\documentclass[twocolumn,showpacs,preprintnumbers,amsmath,amssymb,prb]{revtex4-1}
\usepackage{graphicx}% Include figure files
\usepackage{dcolumn}% Align table columns on decimal point
\usepackage{bm}% bold math
\usepackage{color}
\usepackage{physics}
\usepackage{multirow}
\usepackage[normalem]{ulem}

\newcommand{\rr}{{\bm{r}}}

\def\lsim{\lower.35em\hbox{$\stackrel{\textstyle<}{\textstyle\sim}$}}
\def\gsim{\lower.35em\hbox{$\stackrel{\textstyle>}{\textstyle\sim}$}}

\begin{document}
%\draft

\title{High-temperature superconductivity in flat-band sheared bilayer graphene}

\author{Jos\'e Gonz\'alez$^1$ and Tobias Stauber$^2$ }

\affiliation{
$^{1}$Instituto de Estructura de la Materia, CSIC, E-28006 Madrid, Spain\\
$^{2}$Instituto de Ciencia de Materiales de Madrid, CSIC, E-28049 Madrid, Spain
}
\date{\today}

\begin{abstract}
We propose a new route to induce flat bands with a strong superconducting instability in graphene bilayers with heteroshear, where the 1D character of the moir\'e leads to stronger correlations than in twisted bilayer graphene. We adopt an exact diagonalization approach, on top of a real-space self-consistent Hartree-Fock approximation, to show how the valley polarization of the flat band of a sheared bilayer drives the condensation of Cooper pairs. A unique feature of the 1D moir\'e is that single-particle states with reverse sign of the valley polarization have complementary charge distributions in the moir\'e supercell. This leads to many-body states where the Coulomb repulsion in a Cooper pair is greatly reduced by placing electrons with opposite spin in different valleys. At small hole-doping of the flat band, the many-body ground states are formed by recursive addition of single-hole states, which allows us to reconstruct a quasi-1D Fermi line in the originally flat band. We show that even (odd) numbers of holes lead consistently to ground states with lower (higher) values of the compressibility. This provides the signature of the condensation of Cooper pairs with emergent quasiparticles above a large energy gap, unveiling a strong-coupling route to high-temperature superconductivity in topological flat-band systems.

\end{abstract}

%
% Uncomment for keywords
%\vspace{2pc}
%\noindent{\it Keywords}: XXXXXX, YYYYYYYY, ZZZZZZZZZ
%
% Uncomment for Submitted to journal title message
%\submitto{\JPA}
%
% Uncomment if a separate title page is required
%\maketitle
% 
% For two-column output uncomment the next line and choose [10pt] rather than [12pt] in the \documentclass declaration
%\ioptwocol
%
\maketitle

{\it Introduction.}
The feasibility to engineer moir\'e superlattices by a relative twist in stacked graphene layers has opened a new way to study strong correlation effects in 2D electron systems. The origin of this new trend was the seminal discovery of superconducting behavior next to insulating phases in twisted bilayer graphene (TBG) at the so-called magic angle\cite{Cao18a,Cao18b}. In this setup, it becomes crucial the possibility to form very narrow electron bands by a fine adjustment of the twist angle\cite{Bistritzer}.

\begin{figure}[h]
 \flushleft(a)
 \center
       \includegraphics[width=0.70\columnwidth]{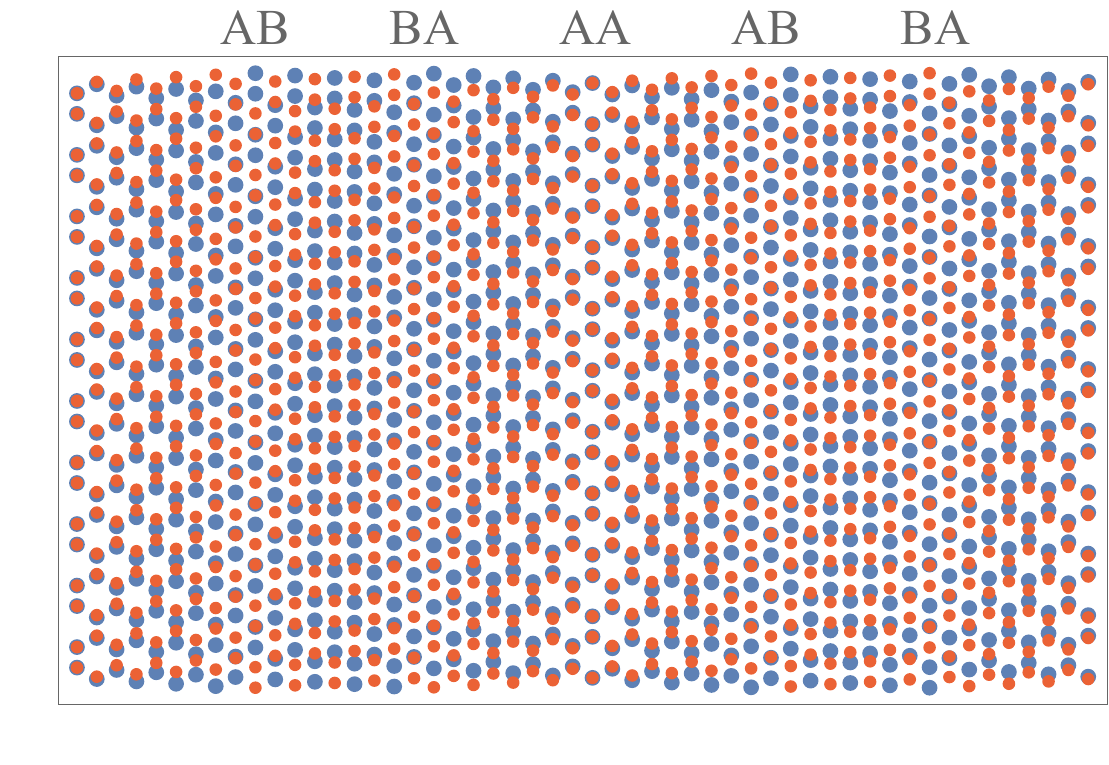}
	\\
 \flushleft{(b) \hspace{3.8cm} (c)}
 \center
	\includegraphics[width=0.49\columnwidth]{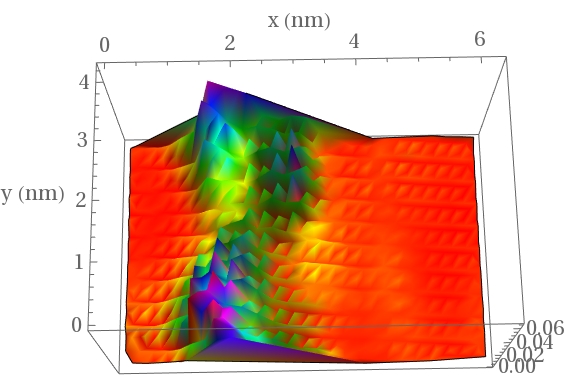}
	\includegraphics[width=0.49\columnwidth]{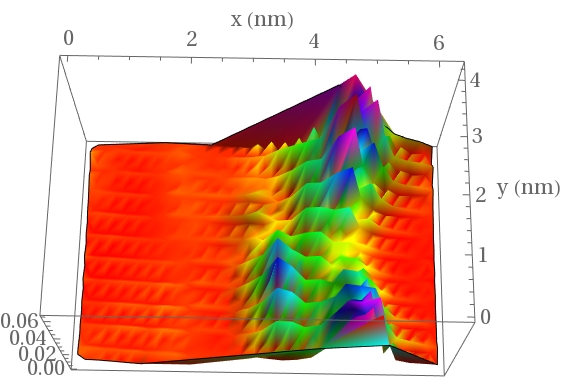}
	\caption{(a) Moir\'e pattern obtained by applying shear to bilayer graphene, showing the formation of an alternating sequence of domains with period $L_x \approx 6.4$ nm in the horizontal direction. Red (blue) dots correspond to the carbon atoms of the upper (lower) graphene layer. (b),(c) Charge distribution of the spinless many-body ground state for filling fraction $2/3$ of the red flat band in Fig. \ref{three}(b), built with single-particle states for negative (b) and positive (c) valley polarization.}
	\label{one}
\end{figure}

In this article we propose a new route to create flat bands and strong superconducting correlations in graphene bilayers introducing heteroshear. This can produce an alternating sequence of regions with $AB, BA$ registry (Bernal stacking) and perfect $AA$ registry between the layers, as seen in Fig. \ref{one}(a). The interacting system is rather unconventional since the 1D character of the moir\'e induces stronger correlations than in TBG. This has a large impact in electronic instabilities like the condensation of Cooper pairs, for which we propose a mechanism leading to a gap with a magnitude only seen in high-temperature superconducting materials.

Domain walls between $AB$ and $BA$ stacking and networks of domains have been observed in bilayer graphene\cite{Brown12,Hattendorf13,Alden13,Butz14,Yin16}, and they are known to support protected 1D states with remarkable transport properties\cite{Jung11,Zhang13,Vaezi13,Li14,Ju15,Barrier24}. We investigate in this paper the effects of the Coulomb interaction, which induces a strong breakdown of valley symmetry in the sheared bilayer. Electrons are then forced to concentrate their charge on one of the domain walls of Fig. \ref{one}(a), with almost a complete suppression of the electron density in the other half of the supercell, as shown in Figs. \ref{one}(b)-(c). As long as this is an spontaneous breakdown of symmetry, electrons with opposite spin projections may choose to have reverse order parameters of valley symmetry breaking. We will show that this leads to a many-body ground state where opposite spins have the charge concentrated on opposite domain walls, as in the two plots of Figs. \ref{one}(b)-(c), with a minimization of the Coulomb repulsion which reinforces the formation of Cooper pairs.

\begin{figure}[htb]
\includegraphics[width=0.47\columnwidth]{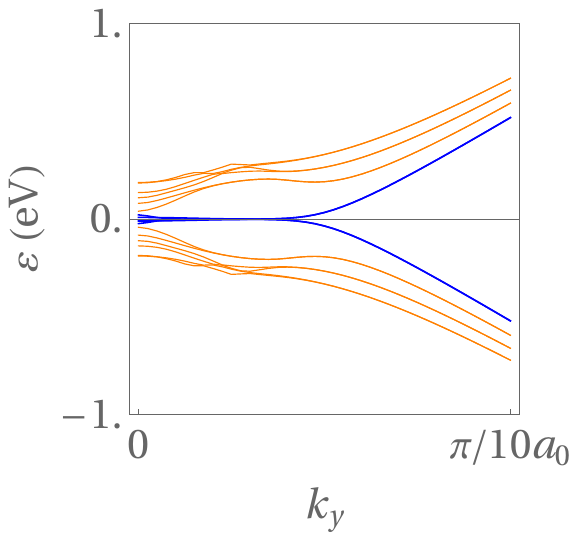}
\includegraphics[width=0.43\columnwidth]{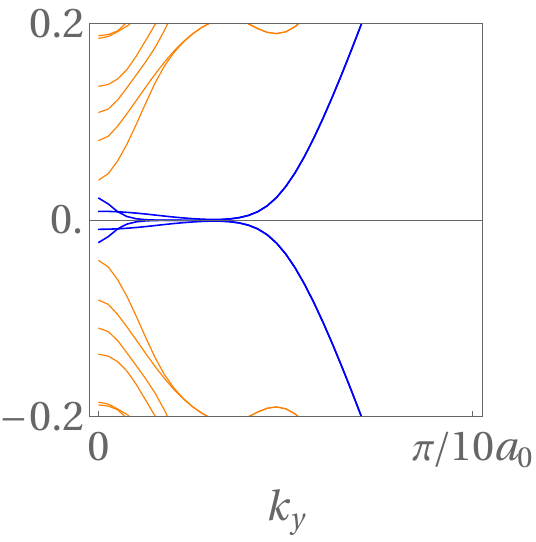}
\\
 \hspace*{1.0cm} (a) \hspace{3.0cm} (b)
\caption{(a) Low-energy bands of a sheared graphene bilayer with period $L_x \approx 56$ nm, obtained from the Hamiltonian of the continuum model with smooth interlayer potentials. $a_0$ is the C-C distance. (b) Zoom view of the flat Landau-like bands in (a).}
\label{two}
\end{figure}

At the theoretical level, it is known that the sheared bilayers have a hidden pseudomagnetic field, which is responsible for the formation of flat bands quite similar to those in the quantum Hall effect\cite{Sanjose12}. This can be best understood in the continuum model, where the Hamiltonian for the four electron amplitudes at $A,B$ and $A',B'$ sublattices of the two layers contains the smooth interlayer potentials $V_{AB'},V_{BA'}$. These can be rewritten in terms of the two components of a non-Abelian gauge field $\hat{\mathbf{A}}$\cite{Sanjose12}. The pseudomagnetic field explains the appearance of flat bands like those shown in Fig. \ref{two}(a). A remarkable difference with respect to TBG is that here, for each valley and spin, four flat bands arise at low energies, a pair of them below and the other pair above the charge neutrality point, as seen in Fig. \ref{two}(b). The flatness of these Landau-like bands is at the origin of the strong correlations arising in the interacting theory.

We will see that, in the phase with strongly broken valley symmetry, the highest occupied band remains quite flat under the effect of the Coulomb interaction. Then, the many-body ground state can be only investigated with methods like the exact diagonalization of the Hamiltonian. At small hole-doping of the flat band, we will show that the ground state has limited entanglement entropy, making possible to reconstruct a quasi-1D Fermi line with strongly renormalized quasiparticles in the originally flat band. The uneven compressibility of the many-body ground states for odd and even filling will allow us then to capture the condensation of Cooper pairs with emergent quasiparticles above a large energy gap, unveiling a strong-coupling route towards high-temperature superconductivity in topological flat-band systems\cite{kopnin11}.

{\it Microscopic Hartree-Fock approach.}
We study the interaction effects by means of a microscopic real-space approach to the sheared bilayer. Our starting point is a tight-binding approximation where the noninteracting Hamiltonian is written in terms of creation (annihilation) operators $a_{i\sigma}^\dagger$ ($a_{i\sigma}$) for electrons at site $i$ with spin $\sigma $
\begin{align}
H_0=-\sum_{i,j} t(\rr_i-\rr_j) a^\dagger_{i\sigma} a_{j\sigma} + \sum_{i} w(\rr_i) a^\dagger_{i\sigma} a_{i\sigma}   \;
\label{h0}
\end{align}

We adopt a dependence of $t(\rr )$ on distance and layer index assuming a usual Slater-Koster parametrization (see Supplemental Material (SM)). At this stage, flat bands across a 2D Brillouin zone can be engineered by taking a periodic potential $w(\rr) = w_0 \sin^2(\pi y/L_y)$, which has the effect of folding the bands of the 1D moir\'e superlattice. 

In the tight-binding approach, the low-energy bands show the same one-parameter scaling operating in the continuum model, which allows us to keep the flat-band regime  by trading larger moir\'e supercells for smaller ones at the expense of reducing the equilibrium interlayer distance $d_\perp^{(0)}$ (see SM). After compression down to $d_\perp^{(0)}/1.18$, we are able to reach the flat-band regime for a not too large moir\'e supercell with 2,080 atoms. This reduction is crucial to carry out a self-consistent Hartree-Fock resolution including all the remote valence bands, which is a must for a sensible description of dynamical symmetry breaking and the low-energy bands in the interacting theory.

In addition to $H_0$, we have the interaction part of the Hamiltonian accounting for the $e$-$e$ repulsion mediated by the Coulomb potential $v(\rr )$,
\begin{align}
H_{\rm int} = \frac{1}{2} \sum_{i,j} a_{i\sigma}^{\dagger}a_{i\sigma} \: v_{\sigma \sigma'} (\rr_i-\rr_j) \: a_{j\sigma'}^{\dagger}a_{j\sigma'}\;.
\label{hint}
\end{align}
We take a spatial dependence of $v(\rr )$ with screening length $\xi = 10$ nm, assuming the presence of nearby metallic gates (see SM). The strength of the Coulomb potential is parametrized by $e^2/4\pi \epsilon$ (in units of eV$\times a_0$, $a_0$ being the C-C distance).

We implement a real-space self-consistent Hartree-Fock approximation by assuming that the interacting electron propagator $G$ can be written in the same way as its noninteracting counterpart $G_0$, but with a set of eigenvalues $\varepsilon_{a\sigma}$ and eigenvectors $\phi_{a\sigma} (\rr_i)$ modified by the interaction\cite{Fetter}. In this notation, $a$ labels the electron band, $\sigma $ stands for the spin projection, and $\rr_i$ runs over the carbon sites of the sheared bilayer. Thus, we have in the static limit
\begin{align}
\left(  G  \right)_{i\sigma,j\sigma} = -\sum_a \frac{1}{\varepsilon_{a\sigma}}  \phi_{a\sigma} (\rr_i)  \phi_{a\sigma} (\rr_j)^*
\label{hf}
\end{align}

The new eigenvalues and eigenvectors are constrained by the Dyson equation
\begin{align}
G^{-1} = G_0^{-1} - \Sigma\;.
\label{dyson}
\end{align}
where $\Sigma $ is the electron self-energy. In the Hartree-Fock approximation, Eq. (\ref{dyson}) provides indeed a closed set of equations, since $\Sigma $ can be written in terms of $\phi_{a\sigma} (\rr_i)$ as
\begin{align}
\left( \Sigma  \right)_{i\sigma,j\sigma}  = &  \; \mathbb{I}_{ij} \:   \sum_{l, \sigma' } v_{\sigma \sigma'} (\rr_i-\rr_l)  \sideset{}{'}\sum_a  \left|\phi_{a\sigma'} (\rr_l)\right|^2      \notag    \\ 
&  - v_{\sigma \sigma} (\rr_i-\rr_j)  \sideset{}{'}\sum_a \phi_{a\sigma} (\rr_i)  \phi_{a\sigma} (\rr_j)^*\;,
\label{selfe}
\end{align}
where $v_{\sigma \sigma'} (\rr )$ stands for the potential of the Coulomb interaction and the prime means that the sum is only over the occupied states\cite{Fetter}.

{\it Spontaneous symmetry breaking.}
We investigate the solutions of Eq. (\ref{dyson}) for different values of the interaction strength $e^2/4\pi \epsilon$, as the dielectric constant $\epsilon $ may vary depending on internal screening as well as on the external environment. As $e^2/4\pi \epsilon$ increases, we observe the onset of new electronic phases with spontaneous symmetry breaking. The Hartree-Fock approximation is well-suited for this task, since the different order parameters can be written in terms of the matrix elements
\begin{align}
h_{ij}^{(\sigma )} =  \sideset{}{'}\sum_a \phi_{a\sigma} (\rr_i) \phi_{a\sigma} (\rr_j)^*\;.
\end{align}

Thus, we have order parameters for the breakdown of time-reversal invariance, which measure the currents around the loops connecting nearest neighbors $i_1, i_2$ and $i_3$ of each atom $i$ in graphene sublattices $A$ and $B$. Two different possibilities can be realized, corresponding to 
\begin{align}
P_{\pm}^{(\sigma )} = & \sum_{i \in A}  {\rm Im} ( h_{i_1 i_2}^{(\sigma )} + h_{i_2 i_3}^{(\sigma )} + h_{i_3 i_1}^{(\sigma )} )  \nonumber    \\
             &  \pm \sum_{i \in B}  {\rm Im} ( h_{i_1 i_2}^{(\sigma )} + h_{i_2 i_3}^{(\sigma )} + h_{i_3 i_1}^{(\sigma )} )  
\label{trsb}
\end{align} 
A nonvanishing $P_+$ is the signature of a Chern insulator phase with Haldane mass\cite{haldane}, while $P_- \neq 0$ signals the imbalance in the energy of the two valleys of the graphene lattice (valley symmetry breaking).

\begin{figure}[h]
\flushleft(a)
 \center
       \includegraphics[width=0.80\columnwidth]{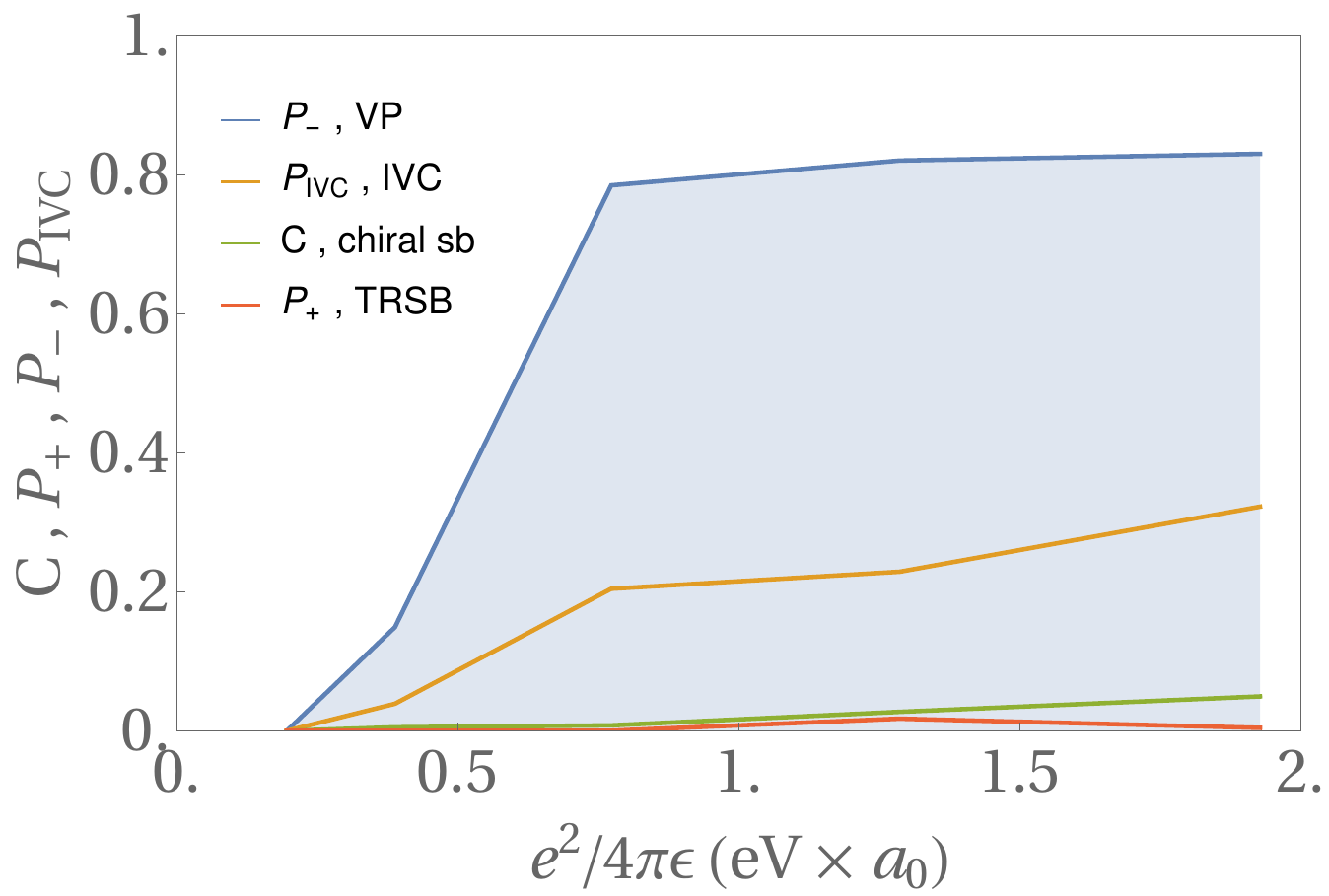}
	\\
 \flushleft(b)
 \center
	\includegraphics[width=0.80\columnwidth]{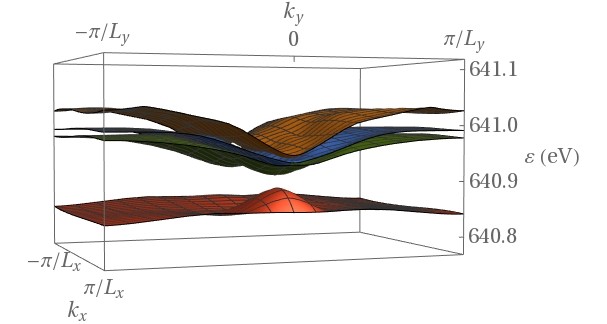}
\caption{(a) Phase diagram of the sheared bilayer with the moir\'e supercell in Fig. \ref{one}(a) and the parameters quoted in the text, obtained with a real-space self-consistent Hartree-Fock approximation for filling fraction $\nu=1$. The order parameters correspond to valley polarization (VP), $K$-intervalley coherence (IVC), chiral symmetry breaking (chiral sb) and time-reversal (parity) symmetry breaking (TRSB). (b) Plot of the lowest-energy conduction bands showing the evolution of the four-fold flat-band manifold above the charge neutrality point after a self-consistent Hartree-Fock resolution for filling fraction $\nu=1$ (flat band in red completely filled) and $e^2/4\pi \epsilon = 1.9$ eV$\times a_0$ ($\epsilon = 5$).}
\label{three}
\end{figure}

Other order parameters like those for $K$-intervalley coherence or chiral symmetry breaking ($P_{IVC}$ and $C$ in Fig. \ref{three}(a)) can be also expressed in terms of $h_{ij}^{(\sigma )}$ (see SM). A systematic approach with their derivation can be found in Ref. \onlinecite{Sanchez24}.

The effects of symmetry breaking become more relevant at integer filling fraction of the flat bands, since then there is a splitting of their degeneracy and the opening of a gap at the Fermi level\cite{Kang19,seo19,Stepanov20,Bultinck20,YiZhang20,Cea20,BiaoLian21,Xie21,Sanchez24}. At filling fraction $\nu=1$ (in the spinless model), there is a critical interaction strength for the onset of valley symmetry breaking, as can be seen in the phase diagram of Fig. \ref{three}(a). Such a symmetry-breaking pattern becomes dominant for all the values of $e^2/4\pi \epsilon$ we have considered, down to $\epsilon = 5$. The imbalance between the two valleys becomes evident in Fig. \ref{three}(b), as the filled red flat band remains below the Fermi level, while the other three bands are above it.

Similar phase diagrams are obtained at filling fraction $\nu=2$ and 3 (spinless model), showing the prevalence of the order parameter for valley symmetry breaking up to the strong-coupling regime (see SM).

{\it Exact diagonalization.} 
After the onset of valley symmetry breaking, some of the low-energy bands may still remain flat in the interacting system, as shown in Fig. \ref{three}(b). Then, a proper determination of the many-body ground state demands the study of correlation effects not captured by the Hartree-Fock approximation. For that purpose, we resort to an exact diagonalization (ED) of the many-body Hamiltonian, built from the single-particle states obtained in the Hartree-Fock approach.

We implement the ED approach in the red flat band shown in Fig. \ref{three}(b), taking grids of $6 \times 4$ and $6 \times 8$ momenta in the rectangular Brillouin zone. These are already sensible discretizations, as the computation of the Chern number from the states in either grid reproduces the correct value $C = 1$ for the band. We first consider the many-body problem in the case of spinless electrons, constructing the Hilbert space of $n$ particles with states 
\begin{align}
| \psi \rangle =  \sum_i \alpha_i  | \chi_i \rangle
\label{hs} 
\end{align}
where the $| \chi_i \rangle $ stand for the basis made of products of single-particle states from the grid in the Brillouin zone
\begin{align}
| \chi_i \rangle = | {\bf k}_{i_1} \rangle | {\bf k}_{i_2} \rangle  \ldots  | {\bf k}_{i_n} \rangle
\label{prod}
\end{align}

\begin{figure}[t!]
\includegraphics[width=0.60\columnwidth]{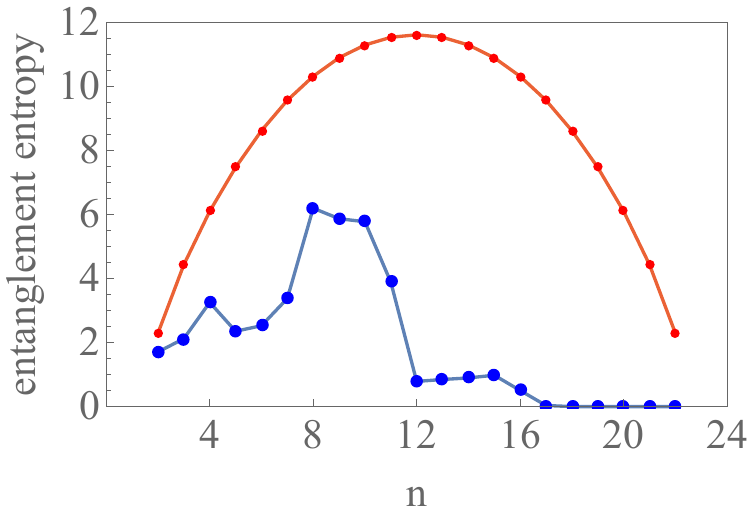}
\includegraphics[width=0.35\columnwidth]{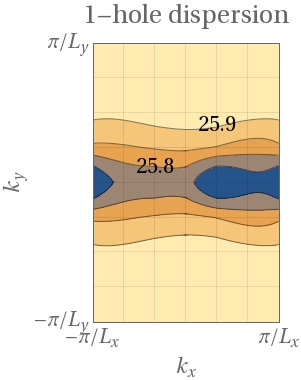}
        \\
	\hspace*{1.5cm} (a) \hspace{3.5cm} (b)
\caption{(a) Entanglement entropy $S$ of the many-body ground states in a $6 \times 4$ momentum grid (blue points) as a function of the number of particles $n$. The red points stand for the maximum possible value of the entanglement entropy, according to the dimension of the space of states. (b) Contour plot of the ground state energy (in eV) for one-hole many-body states in a $6 \times 8$ momentum grid (with $\epsilon = 5$) as a function of the hole momentum $(k_x, k_y)$.}
\label{four}
\end{figure}

In this ED approach, a striking feature is that the ground states we get for small number of holes in the flat band are given essentially by a single contribution (single Slater determinant) in the sum of Eq. (\ref{hs}). The level of entanglement can be quantified by extending the von Neumann expression of the entanglement entropy $S$ to the many-body problem. Then, we define
\begin{align}
	S = - \sum_i | \alpha_i |^2 \log \left( | \alpha_i |^2 \right)  
\label{entr}	
\end{align}
The results for the ground states with different number of particles $n$ are shown in Fig. \ref{four}(a) for the $6 \times 4$ grid (assuming a purely flat band, in the limit of very strong coupling). We observe that the values of $S$ are zero, to very good approximation, for doping levels of up to 7 holes in the flat band. This is in contrast with the large values of $S$ for particle-like doping, reflecting a remarkable particle-hole asymmetry.

The reason for the extremely small values of $S$ at low hole-doping lies in the different pattern of localization of the single-particle states at or near $k_y = 0$ (see SM). These states penalize the energy of a many-body state, already at the Hartree-Fock level, since they have very reduced exchange interaction with the rest of single-particle states. This can be seen from the one-hole dispersion in the ED approach, shown in Fig. \ref{four}(b). From these results, one can actually interpret the reconstruction of a Fermi line enclosing the points $k_y = 0$, since moving the holes away from that line leads to a substantial increase in the energy of the many-body state. At low hole-doping, the ground state is then dictated by the avoidance of single-particle states near $k_y = 0$, as this constrains tightly the content of holes at fixed total momentum.

{\it Cooper-pair condensation.}
When the electron spin is taken into account, there are two different possibilities to assign the two spin projections to the valleys of the sheared bilayer. This is relevant in the phase with valley symmetry breaking, since there is a solution of the self-consistent Hartree-Fock approach in which the two spin projections have opposite values of the valley polarization $P_-$. Looking for the lowest-energy configuration, it is a matter of comparing such a solution to that in which the two spin projections have instead the same value of $P_-$.

An important feature of the spinless ground states of the preceding section is that they have the charge density concentrated mainly on one half of the rectangular moir\'e unit cell, as shown in Figs. \ref{one}(b)-(c). Which half is occupied depends on the sign of $P_-$, since the state with reverse value of the valley polarization has the charge concentrated on the opposite half of the supercell.

It follows then that the appropriate choice to minimize the Coulomb repulsion corresponds to the configuration in which the two spin projections have opposite values of $P_-$. We may approach the many-body ground state by writing the tensor product of two spinless many-body states of the type (\ref{hs}), where the two factors (for spin up and down) correspond to opposite values of $P_-$ with charge distributions like those in Figs. \ref{one}(b)-(c)
\begin{align}
	| \Psi \rangle \approx | \psi_\uparrow \rangle   \otimes   | \psi_\downarrow \rangle
\label{gss}
\end{align}	
The plausibility of this approximation will depend in general on the range of the interaction, since it amounts to neglect the Coulomb repulsion between the almost disjoint charge distributions of the two spin projections.

To check the accuracy of (\ref{gss}) with the ED approach, we double the number of single-particle states in the previous $6 \times 4$ grid, adding now a replica with the states for the reverse values of spin and $P_-$ from the Hartree-Fock resolution. In this model, the dimensions of the space of states are in general exceedingly large, but it is still possible to carry out the ED for small number of holes in the flat band.

We carry out the ED assuming again a purely flat band at very strong coupling. The results show that, up to a maximum of 7 holes for which we have been able to compute (for subspaces of states with dimension $\sim 900,000$), the many-body ground state has in general a dominant component given by a single Slater determinant of the form (\ref{gss}). However, we find now small but significant correlations, contrary to what happened in the spinless model. That is, the many-body ground state cannot be obtained from a simple Hartree-Fock approximation that takes only the diagonal elements in the many-body Hamiltonian. This point is crucial for the existence of pairing correlations, as we see below.

We show in Table \ref{table1} the weight $| \alpha_i |^2$ of the dominant component (\ref{gss}) for each of the ground states depending on the number of particles $n$. For 4 holes ($n=44$) there is a slight mixture with another state of Slater-determinant type, but the important point is that this subdominant component still has a form given by (\ref{gss}). We also show in the Table the single-particle hole states that appear in the dominant component in each case.

\begin{table}[t!]
	\begin{tabular}{c|c|r|l}
		  number of        &   dominant  & hole states   & hole states  \\
		  particles         &   weight    & spin $\uparrow $  & spin $\downarrow $  \\
		\hline \hline
		46  &     0.89         &  (-2,0)  &  (2,0)                \\
		\hline
		45   &   0.89          &  (1,0),(-2,0)  &  (2,0)    \\
		\hline
		44  &    0.72+0.17     &  (1,0),(-2,0)  &  (2,0),(-1,0)         \\
		\hline
		43  &     0.86         &  (3,0),(1,0),(-2,0)  &  (2,0),(-1,0)      \\
		\hline
		42  &     0.82         &  (3,0),(1,0),(-2,0) &  (2,0),(-1,0),(-3,0)      \\                
		\hline 
		41  &     0.87         &  (2,0),(3,0),(1,0),(-2,0) &  (2,0),(-1,0),(-3,0)      \\                  
		\hline
	\end{tabular}
	\caption{Table showing the weight $| \alpha_i |^2$ of the dominant component (\ref{gss}) in the many-body ground states obtained from ED at low hole doping of a $6 \times 4$ momentum grid. We quote also the single-particle hole states in such a dominant component, labeled according to their position $(i_x,i_y)$ in the grid (with $-3 < i_x \leq 3$, $-2 < i_y \leq 2$).} 
	\label{table1}
\end{table}

The remarkable property of the many-body ground states we obtain is that they are formed, to first approximation, by a product of Cooper pairs, as seen in Table \ref{table1}. This is a consequence of the fact that the single-particle states we find in the spin down factor $\psi_\downarrow $ are the partners by valley (inversion) symmetry of the single-particle states in $\psi_\uparrow $. This property, which holds for the ground states, cannot be taken for granted, however, since it does not apply in general to the many-body excited states.

As seen from Table \ref{table1}, the $n$-th ground state can be obtained, to good approximation, by adding a single-hole state to the ensemble of single-particle states of the $n+1$ ground state in the table. The ground states for odd $n$ can be viewed then as quasiparticle states for the Cooper-pair condensates from which they arise.

\begin{figure}[t!]
\includegraphics[width=0.70\columnwidth]{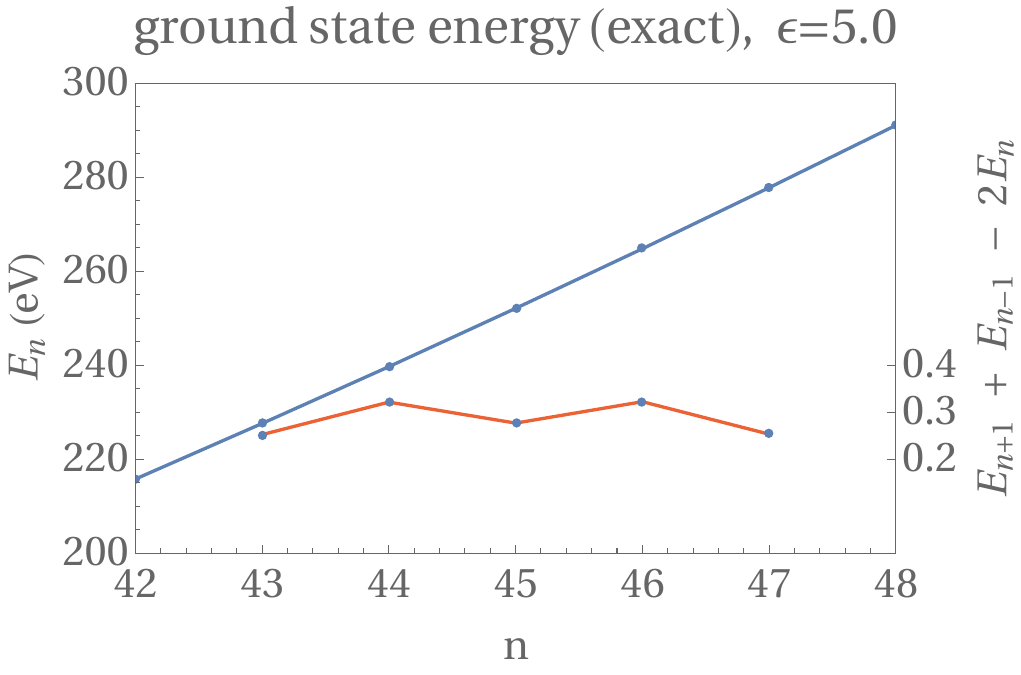}
\caption{Plot of the energy $E_n$ of the many-body ground states as a function of the number of particles $n$, at low hole-doping of the $6 \times 4$ grid and $\epsilon = 5$ (blue curve). The red line joins the points of the curvature $E_{n+1} + E_{n-1} - 2E_n$, measured with the scale to the right of the figure (in eV).}
\label{five}
\end{figure}

The phenomenon of Cooper-pair condensation is validated by the results for the ground state energy $E_n$ as a function of $n$, represented in Fig. \ref{five}. The plot shows that the points with even $n$ have in general greater curvature, given here by 
\begin{align}
\kappa_n = E_{n+1} + E_{n-1} - 2E_n 
\end{align}
That means also greater stability, as it implies that the ground states with odd $n$ have an excess of energy over the curve for even $n$. We stress that this effect arises from the correlations in the many-body ground states, since the ground-state energies in the Hartree-Fock approximation do not show any sign of a similar odd-even mismatch in the values of the curvature (see SM).

By drawing two different ground-state energy curves for odd and even $n$, we can estimate the gap $\Delta_n $ as the additional energy needed to create a quasiparticle in the ensemble of Cooper pairs (see SM). Thus, we get 
\begin{align}
\Delta_{46} & \approx 12 \; {\rm meV}       \\
\Delta_{44} & \approx 11 \; {\rm meV}
\label{gap}
\end{align}
which offers good prospects to observe the strong-coupling condensation in the flat band of the sheared bilayer.

{\it Conclusion.}
We have seen that the sheared graphene bilayers can be tuned to have flat low-energy bands for sufficiently large size of the moir\'e supercell. In that regime, the interacting system becomes prone to develop broken-symmetry phases, with valley symmetry breaking as the dominant pattern. This is a main distinction with respect to TBG, where there is a natural competition between valley symmetry breaking and intervalley coherence, specially at strong coupling.  

The strong signal of valley symmetry breaking favors the onset of a pairing instability in which the electrons with opposite spin projection in the Cooper pairs live in different valleys. The origin of this effect lies in the complementary charge distributions of the states attached to opposite valleys, which favor the reduction of the Coulomb repulsion.

Although we deal with a strong-coupling approach, we can still draw a connection with the conventional BCS theory of superconductivity. Superconducting instabilities are usually discussed within the Bethe-Salpeter equation in the Cooper channel. Irrespective of the pairing mechanism, the critical temperature is then given by $k_BT_c=\Lambda_ce^{-1/\lambda}$, where $\Lambda_c$ denotes the energy scale of the pairing mechanism and $\lambda$ is the pairing strength. In the case of a purely electronic Kohn-Luttinger mechanism, the superconducting instability arises from the distortion of the Fermi lines, which leads to anisotropic screening and attraction in some of the interaction channels\cite{Kohn65,Baranov92}. $\Lambda_c$ is then related to the bandwidth and $\lambda$ corresponds to the absolute value of the most negative eigenvalue of the Cooper-pair vertex.

In our strong-coupling approach, the critical temperature is also limited by the overall energy scale, i.e., $k_B T_c\leq\Lambda_c$. Since our starting point for the sheared bilayer is an extremely narrow band, the critical temperature could be expected to be rather small. However, the {\it effective} bandwidth of the correlated quasiparticles is strongly renormalized, yielding $\Lambda_c \sim 0.1$ eV (see Fig. \ref{four}(b)). We further note that the Kohn-Luttinger mechanism is usually stronger for anisotropic band structures. It seems therefore plausible to set $\lambda \sim 0.5$, which was also numerically calculated in Ref. \onlinecite{Gonzalez19} for TBG. This gives the order of magnitude of $k_BT_c \sim 10$ meV, consistent with the above estimate in the strong-coupling approach.

A similar mechanism may operate in other moir\'e systems like TBG, which could explain some distinctive experimental features like the hole-doped regime of the superconductivity or the strong binding of the Cooper pairs. Anyhow, the sheared bilayer has the natural effect of enhancing the electron correlations in the 1D moir\'e pattern. This reflects in the magnitude we obtain for the gap in the Cooper-pair condensate, much larger than that of TBG, and in the range of materials exhibiting high-$T_c$ superconductivity.

{\it Acknowledgements.}
The work was supported by grant PID2023-146461NB-I00 funded by MCIN/AEI/10.13039/501100011033 as well as by the CSIC Research Platform on Quantum Technologies PTI-001. The access to computational resources of CESGA (Centro de Supercomputaci\'on de Galicia) is also gratefully acknowledged.

\end{document}